\documentclass[conference]{IEEEtran}
\usepackage{makeidx}  % allows for indexgeneration
\usepackage{graphicx} \usepackage{epsfig,epsf} \usepackage{subfigure}
\usepackage{enumerate} \usepackage{balance} \usepackage{amsmath}
\usepackage{listings}
\usepackage{color}
\usepackage{booktabs}
\usepackage{balance}
\ifCLASSINFOpdf
\else
\fi

\begin{document}

\title{OSCAR: Object Security Architecture for the Internet of Things}

\author{
	\IEEEauthorblockN{
		Mali\v{s}a Vu\v{c}ini\'{c}\IEEEauthorrefmark{1}\IEEEauthorrefmark{5},
		Bernard Tourancheau\IEEEauthorrefmark{1},
		Franck Rousseau\IEEEauthorrefmark{1},
		Andrzej Duda\IEEEauthorrefmark{1},
		Laurent Damon\IEEEauthorrefmark{5},
		Roberto Guizzetti\IEEEauthorrefmark{5}
	}
	
	\IEEEauthorblockA{ 
		\IEEEauthorrefmark{1}Grenoble Alps University, CNRS Grenoble Informatics Laboratory UMR 5217, France
	}
	
	\IEEEauthorblockA{
		\IEEEauthorrefmark{5}STMicroelectronics, Crolles, France\\
		Email: \{firstname.lastname\}@imag.fr, \{firstname.lastname\}@st.com
	}
}

% make the title area
\maketitle

\begin{abstract} Billions of smart, but constrained objects wirelessly connected
  to the global network require novel paradigms in network design. New
  protocol standards, tailored to constrained devices, have been designed taking
  into account requirements such as asynchronous application traffic, need for
  caching, and group communication. The existing connection-oriented security
  architecture is not able to keep up---first, in terms of the supported
  features, but also in terms of the scale and resulting latency on small
  constrained devices. In this paper, we propose an architecture that leverages
  the security concepts both from content-centric and traditional
  connection-oriented approaches. We rely on secure channels established by
  means of (D)TLS for key exchange, but we get rid of the notion of the
  ``state'' among communicating entities. We provide a mechanism to protect from
  replay attacks by coupling our scheme with the CoAP application protocol. Our
  object-based security architecture (OSCAR) intrinsically supports caching and
  multicast, and does not affect the radio duty-cycling operation of constrained
  objects. We evaluate OSCAR in two cases: 802.15.4 Low Power and Lossy
  Networks (LLN) and Machine-to-Machine (M2M) communication for two different
  hardware platforms and MAC layers on a real testbed and using the Cooja
  emulator. We show significant energy savings at constrained servers and
  reasonable delays. We also discuss the applicability of OSCAR to Smart City
  deployments.
\end{abstract}

\section{Introduction}
\label{introduction}
% General introduction discussing IoT, application scenarios (including type of information exchanged - sensor readings, actuating commands, etc. ), criticality of securing the network from both, user's privacy and operating perspective (due to the constraints). State that we should decouple protection of the information and protection of the network itself and add a discussion on current trust model in the Internet (information trust relies on the authenticity of the host). Briefly state the idea of adding security overhead within the  payload to protect the information. Briefly state the contributions in sense of surveying the work in different areas of Computer Science and discussing its applicability for IoT.
The long awaited Internet of Things (IoT) has never been closer. The upper layers of the IP protocol stack for constrained devices are being fine-shaped and the gaps between IETF and IEEE standards are being bridged. Security remains a concern. Well designed IP security protocol suites have been ported to constrained devices of IoT. Their core design assumptions, however, build upon the connection-oriented trust model that poorly fits the IoT requirements. Nevertheless, past experiences have shown that designing a security protocol right is a tough and error-prone process. 

Research efforts towards the secure IoT have thus mostly concerned designing
lightweight variants and porting them to constrained devices \cite{sizzle,
  lithe, raza2011securing, raza20126lowpan}, which led to a situation where the
nearly standardized Constrained Application Protocol (CoAP) \cite{coap-draft}
fully supports the application requirements, but security does not keep
up. Smart devices, due to their severe energy and memory constraints, heavily
rely on group communication, asynchronous traffic, and caching. Supporting a
variety of existing security protocols/mechanisms to specifically target these
requirements is practically impossible due to memory limitations. IETF has thus
taken a position \cite{coap-draft} to reuse Datagram Transport Layer Security (DTLS),
 the all-round point-to-point security protocol, to secure the communication
 channel between a constrained node running a CoAP server and a client.

Apart from the inherent incompatibility with multicast traffic and caching, the
plain DTLS approach has an important impact on scalability. Namely, memory
limitations of constrained servers restrict the total number of handled security
sessions. In IoT scenarios, such as smart city deployments, where a large number
of clients per constrained server is expected, the limitations lead to a
considerable load on the server to handle security associations with each
client. The load translates into increased energy consumption and a shortened lifetime of devices. 

We address this problem from a networking perspective and follow the Representational
State Transfer (REST)
architecture model \cite{rest} to remove the notion of the state between a
server and a client even in terms of security. We achieve this by leveraging the
concept of object security that protects the information content itself. We
couple the object security principles with the capability-based access control
to provide communication confidentiality and protect from replay attacks. Yet,
we fully leverage a vast amount of work behind the (D)TLS protocol and use
secure channels for authenticated key distribution.

The main contributions of the paper are the following:
\begin{itemize}
\item a new scalable security architecture for IoT that jointly provides end-to-end security (E2E) and access control, decouples confidentiality and authenticity trust domains, and intrinsically supports multicast, asynchronous traffic, and caching, 
\item an evaluation of the architecture in a constrained Machine-to-Machine (M2M) scenario for two
  hardware platforms and MAC layers, on a real testbed and in the instruction level emulator of Cooja, demonstrating performance benefits with an increasing number of clients.
%\item an open source object security software library tailored for the Contiki
%  operating system and constrained devices.
\end{itemize}

The paper is organized as follows. We discuss the current Internet trust model and the requirements of  IoT applications in Section \ref{internet-trust-model}. We provide a detailed description of the proposed architecture in Section \ref{oscar}, discuss security considerations in Section \ref{security},
and evaluate it in Section \ref{evaluation}. Section \ref{relatedwork} summarizes the
related work. We conclude and discuss the future work in Section \ref{conclusion}.
 
 \section{Internet Trust Model and the IoT Requirements}
\label{internet-trust-model}
% General discussions from Content Centric Networking on how the security in the Internet was added later on, and how the trust in the Internet comes from the authenticity of the host, and not of the information. 
The fact that the Internet had been designed to facilitate host to host communication has had direct repercussions on the security design. Namely, security followed the model by placing the trust on end points and securing the communication channel. As applications evolved, the Internet has become a content distribution network leveraging the legacy client-server architecture.  This paradigm has led to substantial research efforts on future Internet architectures \cite{ccn, dona}. 
%One of the promising directions are information centric networks, such as DONA \cite{dona} and Content-Centric Networking \cite{ccn}. 
Our work leverages their contributions and applies the general concepts with the goal to provide a robust, but flexible security approach to IoT and its traffic requirements. %In essence, our approach keeps secure communication channels open for key distribution, but removes notion of state among communicating entities.

As discussed by Smetters and Jacobson \cite{securing-content}, the host oriented
paradigm has a direct consequence on trust---its transitivity: once a
logical connection between the hosts is closed, the trust in the information is
gone. The model serves very well typical e-commerce, e-banking, or IP telephony
applications, because the trust in the information is implicitly dependent on
the trust of the communicating entities \textit{during the connection time}. 
 
The difficulty arises once the notion of a connection disappears. As stressed by
Modadugu and Rescorla~\cite{dtls}, DNS is purposely secured with the application
level extension DNSSEC and not with a connection-oriented protocol, such as DTLS.  Electronic mail, passing multiple application level gateways and without clear connection between end points, is secured with S/MIME or PGP. Applications encompassing IoT emerge as another example, because:
 
 \begin{itemize}
 \item \textit{Application traffic is asynchronous.} Servers (event detectors, monitoring sensors, smart meters) notify their clients (subscribers) of physical state changes as they happen. Clients send commands to actuating devices asynchronously as the changes in the environment are observed. DNS traffic is a good parallel as it is triggered by  asynchronous human actions. 
 \item \textit{Caching is a must.} Severe energy constraints lead to servers
   being asleep more than 99\% of the time. As an already supported (without
   security) and intuitive mechanism, caching at untrusted intermediaries is a
   way to keep applications running independently. A similar problem is faced with electronic mails, as they are stored at untrusted servers until delivery.
 \item \textit{Group communication is frequent.} Commonly, clients instruct a subset of all devices to perform an action, for example to turn off all lights on $n^{th}$ floor or to update the firmware. To achieve this, IPv6 multicast and UDP are exploited bearing no connection state between end points.
 % [Extended version]
 %\item \textit{Servers have severe limitations on non-persistent memory.} State preservation at any level of the protocol stack leads to simple DoS attacks.
 % [Extended version]
 %\item \textit{Connection-less, unreliable transport is generally preferred.} Reliability is left as an option to applications, on a per-need basis.
 \end{itemize}

Typical Web applications are built around a single logical server and multiple
clients \cite{rest}. As a consequence, access control is often done within the server side application, once the client has been authenticated. % [MV] CHECK THIS, NOT COMPLETELY SURE!
IoT reverses this paradigm by having many devices serving as servers and
possibly many clients, taking part in the same application. More importantly,
servers are significantly resource constrained, which results in the
minimization of the server side functionality. Subsequently, access control
becomes a distributed problem, especially when taking into account the recent efforts of decoupling the sensor network infrastructure from applications \cite{senshare, m2m-metering}. Furthermore, applications have emerged that use local databases to store parts of collected data \cite{database}.
 
Recognizing these requirements, it is clear that the connection-oriented trust
model is not the best fit for the actual needs of IoT. It is
true that with different sorts of connection time tweaking and keep-alive
messages we could squeeze in connection-oriented security protocols and work
around the asynchronous traffic requirement. Aside the overhead, this would
still provide us only with the communication channel security. To support
caching, we would need to trust the intermediate nodes/proxies to store the
data. Note that we deal with devices physically accessible to anyone. To support
group communications, we would need to open separate secure connections among
group members and/or add additional protocols on top of them, which effectively
provides redundant security services necessary for use cases. Such a solution is
not a long term approach.
 
Nevertheless, we do not argue that we should ditch well studied
connection-oriented security protocols from the IoT picture. In fact, OSCAR
relies on secure and authenticated channels established by means of DTLS for key
distribution: our approach brings together the concepts of
connection-oriented security with those of content-centric networking
\cite{ccn}.

\section{OSCAR} %Some catchy name needed here
\label{oscar}
We argue that we can meet the discussed requirements by leveraging the benefits of the ``object security'' concept. At the same time, we can provide much greater flexibility to the system as a whole. 

\subsection{Technological Trends and Design Goals}

\label{trends}
Future trends are hard to predict, but a decade of research on Wireless Sensor Networks has given us a lot of insights. Accordingly, we draw the following conclusions that guide our design:
\begin{itemize}
\item Constraints on energy are almost constant. Without a breakthrough in
  chemical engineering, the available energy is expected to remain the main constraint for IoT devices. 
\item Available memory for embedded devices slowly increases. However, due to
  the economical and energy cost caused by leakage in SoC, we expect that memory
  will remain limited and a determining factor for the unit price. 
\item Processing capabilities constantly increase even for ultra low power micro
  controllers. Thus, we do not see the processing power as a limiting constraint in the future.
\end{itemize}

Apart of sleep mode leakages, the energy consumption is mainly caused by radio
communications. Thus, our primary design goal is to minimize the number of extra
frames/packets that need to be transmitted or received for pure security
purposes. We achieve this goal by leveraging the benefits of public key
cryptography, sparse traffic patterns within local constrained networks, and
messages of a limited size---we trade the radio usage for a higher computation
load.

\subsection{Producer-Consumer Model}
\label{producer-consumer-model}
We can abstract IoT, its sensors and actuators, as an interface to the physical
world. Decision takers (human users, intelligence centers, or constrained
actuating devices themselves) base their reasoning on input data coming from the sensed physical phenomena. The relation between enforced decisions and sensed phenomena is \textit{many to many}---a single measurement often affects multiple decisions and a single decision may be based on many different phenomena.
Consider for instance a traffic control application in a Smart City.  A traffic
light management subsystem may use the current traffic intensity and pollution
readings from all over the city as input data for control decisions. At the same
time, local readings may influence decisions made on luminosity of nearby street
lights. 

We believe that the producer-consumer model represents well our problem
also in terms of security. Producers (smart meters, traditional sensors, motion
detectors, etc.) feed consumers with the required information. Consumers
(actuating devices, collection centers, human users) gather up the information
and may further \textit{generate} actions. Actually, the inspiration for the use
of the model comes from Cloud Computing and a recent work by Jung~\emph{et al.}
on data access control \cite{jungprivacy}. An important difference, however, is
that producers in the IoT case are not access control decision makers for the
content they generate, which is rather a policy of the network operator. 

Producers should, thus, care about generating and securing the content or in the
REST terminology, the resource representations, and not about
consumers. Consumers with appropriate access 
privileges should make sure they can make use of the fetched content by
decrypting and authenticating its validity.

% Not well placed, add this as an example once the technical proposition is outlined. 
%To illustrate the applicability of the concept, imagine a simple scenario where an application requests current temperature readings from around the building in order to adjust the HVAC system. Each device then ``produces'' its response by formatting the current reading, and possibly adding meta-data, such as a timestamp and application related information. The content is then signed/authenticated with device/group specific credentials (certificate, symmetric key - depending on the processing capabilities).  Before responding to the request, an application protocol, such as CoAP or HTTP, fetches a secret associated with the temperature resource. Then, content and its signature are encrypted using a key derived from the resource-specific secret. Note that we separately treat two commonly interleaved security services, in a way that:

Following the discussion, we believe that the extent of security tasks performed
by producers should be minimized---producers should not waste precious
resources on exchanging security handshake messages with each consumer. 
From the producer perspective, there are two main reasons for this situation:
\begin{itemize}

\item \textit{Resource representations are minimal in size.} The generated
  content, \emph{i.e.}, the resource representations are typically the measurements of
  physical quantities or different states of a device with possibly additional
  information such as location and a timestamp, which very often makes them smaller in size than individual messages exchanged during a security handshake.  As a consequence, responding with an access-protected resource representation is cheaper than performing multiple RTT handshakes.
\item \textit{Due to the physical constraints, the number of supported
    cryptographic ciphers is limited.} Indeed, constrained devices often have a
  single supported cipher suite (selected at the compile time). This fact
  reverses the paradigm encountered in the Internet where one of the security
  concerns during the handshake is the downgrade attack (the attacker forces two
  parties to use the weakest common cipher by altering the set supported by the
  client). The motivation behind the attack is the assumption that the client
  cipher set is just a subset of those supported by a resource rich server. With the reversed paradigm, as in the IoT case, the motivation for the attack fades away.

% [Extended version] -- not also number of reasons above
%\item \textit{Client-side authentication is as important as server-side.} This has already been discussed but the experimental results show the in-feasibility of the DTLS approach due to the completion delays, even with very powerful embedded devices \cite{kothmayr2012dtls}. Client authentication within the application would inevitably result in additional exchanges leading to unnecessary energy expenditure at server side and is usually out of the question for M2M deployments.  
\end{itemize}

One of our goals is to offload the burden of authentication from constrained servers and to place it on more powerful devices. Such semi-trusted third parties could be physically secured nodes in the network and/or hosts in the Cloud. Their role would be to authenticate individual consumers and share with them appropriate \textit{access secrets} and necessary certificates. We define an ``access~secret'' as an access token from which symmetric encryption keys are derived. Later, consumers can fetch the protected content either from intermediate proxies or directly from producers.
% Missing some terminology on access tokens, etc... 
%Maybe add an explanation that we do not negotiate ciphers to be used, rather that it is the consumer's responsibility to be able to decrypt the content.

%AD: up to here...

\begin{figure}[htbp]
\centering
\includegraphics[width=1\columnwidth]{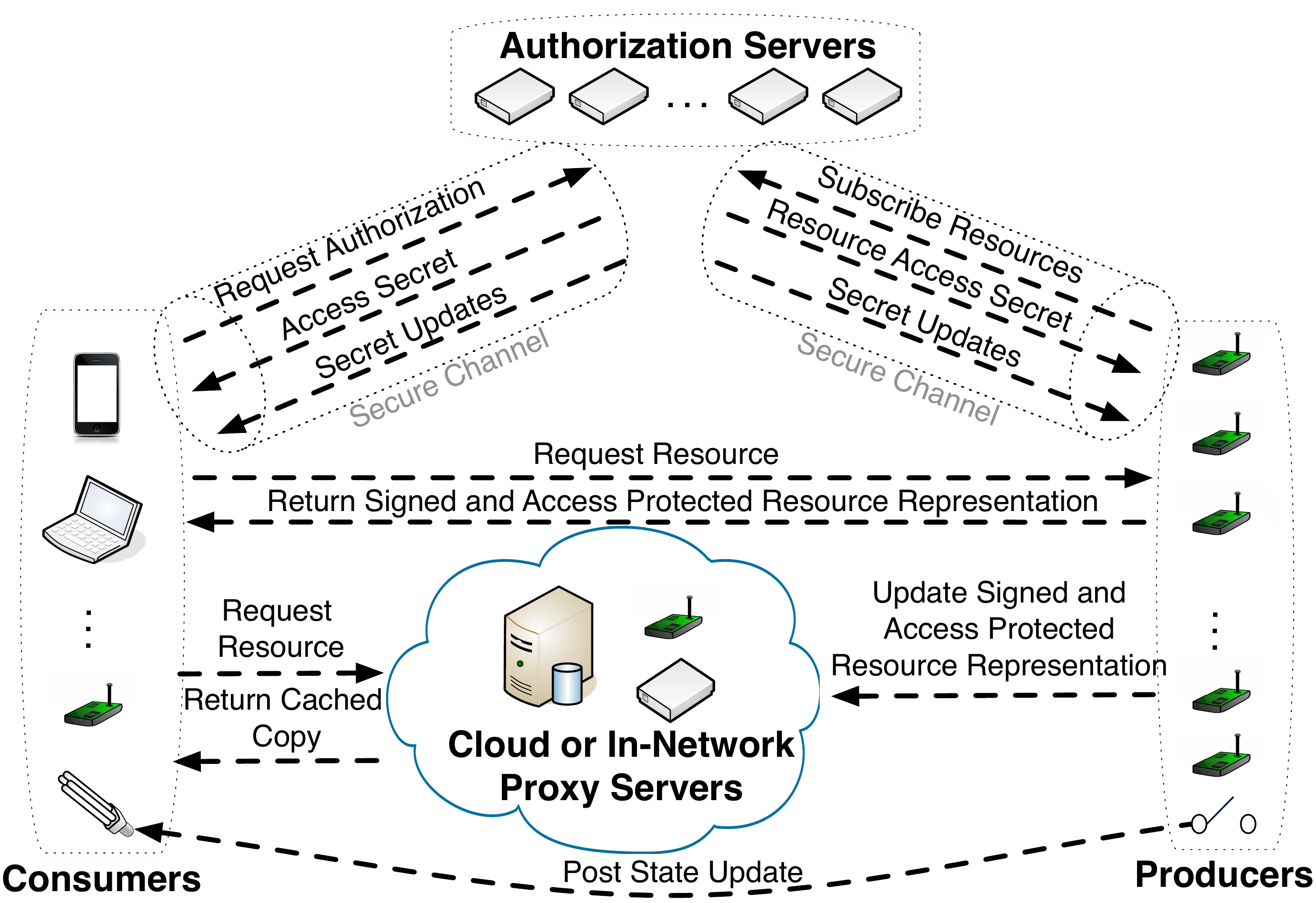}
\caption{OSCAR, a producer-consumer security model for IoT based on object security.}
\label{fig:system-model}
\end{figure}

We thus consider separately \textit{in terms of trust} two commonly interleaved security services:
\begin{itemize}
\item \textit{Confidentiality is used as a means to provide capability-based access control and a protection against eavesdropping.} As a consequence, third parties in charge of authorization need to be trusted. 
\item \textit{Authenticity and integrity of content is tied to the host.} 
Consumers independently decide if they will trust the source \textit{for the provided content}. For example, temperature readings would need to be signed by a server that is certified to have a temperature sensor and to be deployed in the wanted physical location.
\end{itemize}

%AD: still does not read well - how one can give away confidentiality?
%MV changed 'give away' to 'disclose', it should make more sense now.
This approach could be interpreted as we disclose true E2E confidentiality 
to third parties. 
However, even in the classical TLS scenario,
the authorization authority running on a server has potential access to all the
information flowing on the secured channel. %As in the IoT case, the
%authorization authorities are separated from constrained servers, we are
%AD: which concept?
%essentially extending the concept.
%Therefore, we essentially keep the same level of confidentiality 
%We discuss this issue in more details in
%AD: forward reference?
%Section \ref{security}.
Note that authenticity and integrity of information are
not affected and the traditional E2E properties are preserved.
%AD: There is no description of this figure - what is going on if you
%communicate with a device.
%MV: Direct communication is explained in Fig. {fig:client-server}, in more details.
% I added a clarification here though. 

Fig.~\ref{fig:system-model} illustrates the abstract model and the logical
interactions. Notice that from the perspective of a producer, a direct request-response
interaction with a consumer is handled equally as in the case of asynchronous 
updates. More precisely, producers locally keep cached and secured
resource representations, \emph{i.e.} signed objects, and use them to feed different types of consumers.
% To make things concrete, we discuss the implementation of the concept using
% recently standardized CoAP protocol.
%AD: I would suggest "association" instead of "connection".
%MV: Yes, fits better.
In essence, we remove any notion of a logical association between a producer and
a consumer.

In the following, we assume that producers and consumers already possess necessary certificates. 
% [Extended version]
%In Section~\ref{bootstrapping}, we discuss a simple way how this could be bootstrapped by leveraging the object security concept and the scheme in Figure \ref{fig:system-model}. 
During the resource subscription phase, producers publish their certificates to
Authorization Servers. If a certain certificate is required for signature
verification purposes, consumers can fetch it from them. In this way, we remove the burden of certificate transmission from constrained servers to their multiple clients.
% [Extended version]
%Note that due to the certificate sizes, packets containing them are often fragmented --- resulting in multiple frames in case of low power and lossy link layers. This imposes significant load on servers as they would need to do it separately for each client.

\subsection{Fitting the Concept with the REST Architecture and CoAP}
\label{coupling-with-coap}
While object security is traditionally used by applications themselves, we
discuss here the coupling with CoAP, the RESTful application protocol. By doing
%AD: how REST can provide security?
so, we aim to make a bound between secured objects and the underlying communication
protocol, in order to protect against network adversaries in a stateless manner.
Note that in the rest of the paper, we use the REST terminology and
refer to producers as REST servers and consumers as REST clients. 
%AD: strange - you have a CoAP client and a server on a sensor node?
%MV: Yes, it's fairly typical. Consider motion detector and a wireless switch on the same device.
In IoT deployments, however, the same physical device often plays both roles.

% \begin{table}[htbp]
% \centering
%     \caption{Notation \label{tab:traffic-patern}}
%     \begin{tabular}{ c p{6cm}}
%     \toprule
%     $R_i$ & $i^{th}$ publicly available resource \\ %\hline
%     $S_j$ & $j^{th}$ access-secret resource \\ %\hline
%     $A_k$ & $k^{th}$ trusted authority \\ %\hline
%     $Sign(A, x)$ & Digital signature of $x$ with $A$'s private key\\ %\hline
%     $Encrypt(S, x)$ & Symmetric encryption of $x$ with key derived from $S$ \\ %\hline
%     \bottomrule
%     \end{tabular}
% \end{table}

\begin{figure}[htbp]
\centering
\includegraphics[width=1\columnwidth]{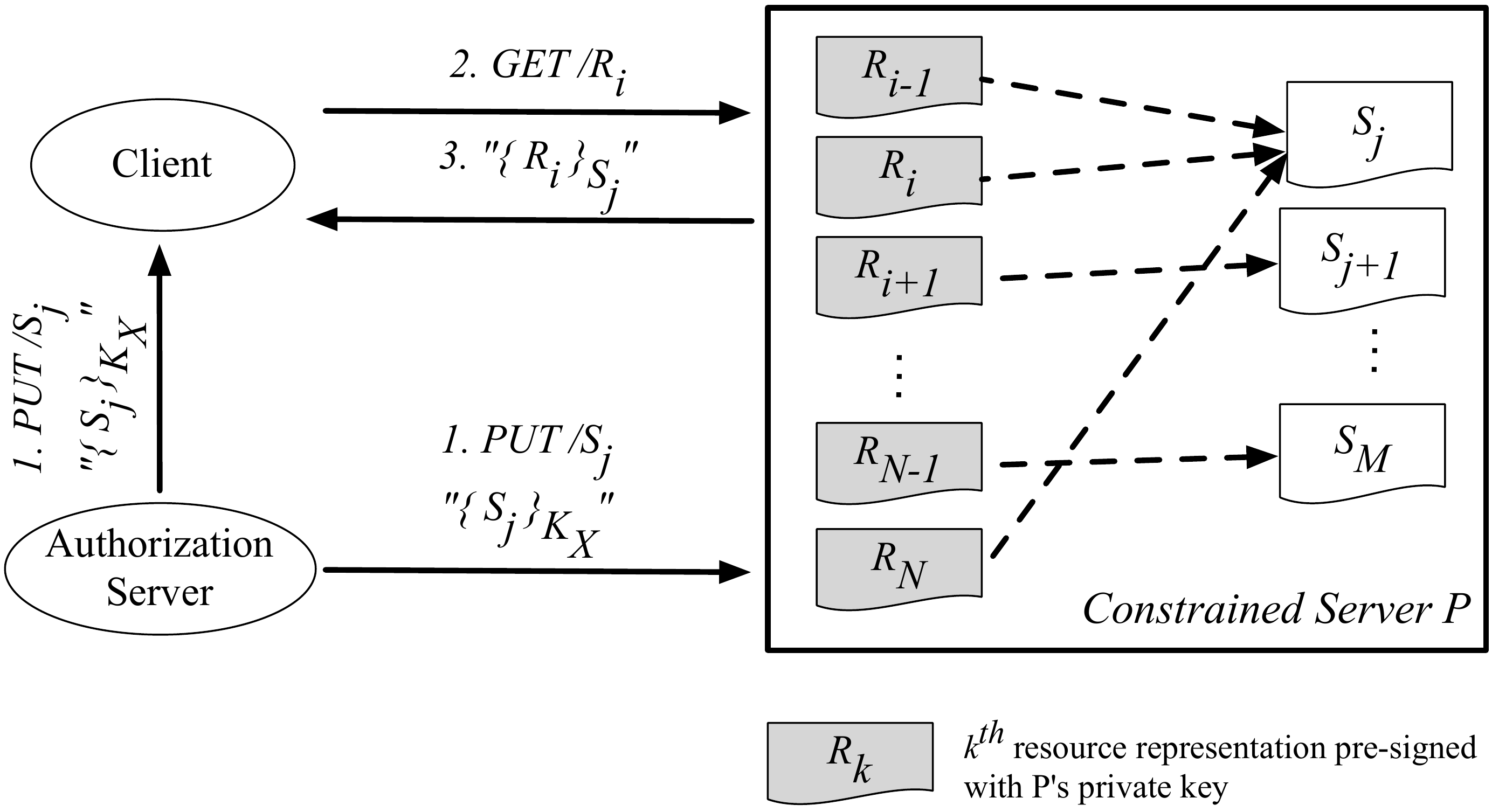}
\caption{Coupling OSCAR with the RESTful architecture of CoAP. Note that
  server-side digital signing operations are performed offline.  $\{X\}_K$
  denotes the online symmetric encryption of $X$ with the key derived from
  $K$. Encryption key $K_X$ of a new access secret  may depend on the key management scheme.% $A_k$, $Sign(A, x)$, $Encrypt(S, x)$ are respectively the $k^{th}$ trusted authority, the digital signature of $x$ with $A$'s private key, the symmetric encryption of $x$ with the key derived from access secret $S$.
}
\label{fig:client-server}
\end{figure}

We abstract the access secrets as REST resources, which allows using the
idempotent PUT method to create or update them. Servers, then, allow the change
only if the enclosed object, \emph{i.e.} a new access secret, has been signed
by the trusted authority. The relation
between $j^{th}$ access secret $S_j$ and $i^{th}$ resource $R_i$ is dependent
on authorization policies and the desired level of confidentiality. It is a
part of resource $R_i$ itself allowing for different confidentiality and
access right resource groups (cf. Figure \ref{fig:client-server}).
%Note that the actual encryption key is locally derived from the secret,
%allowing use of different ciphers. 

In the following, we discuss two important aspects from the communication point of view. 
\subsubsection{Replay protection} 
\label{replay}

\begin{figure}[htbp]
\centering
\includegraphics[width=1\columnwidth]{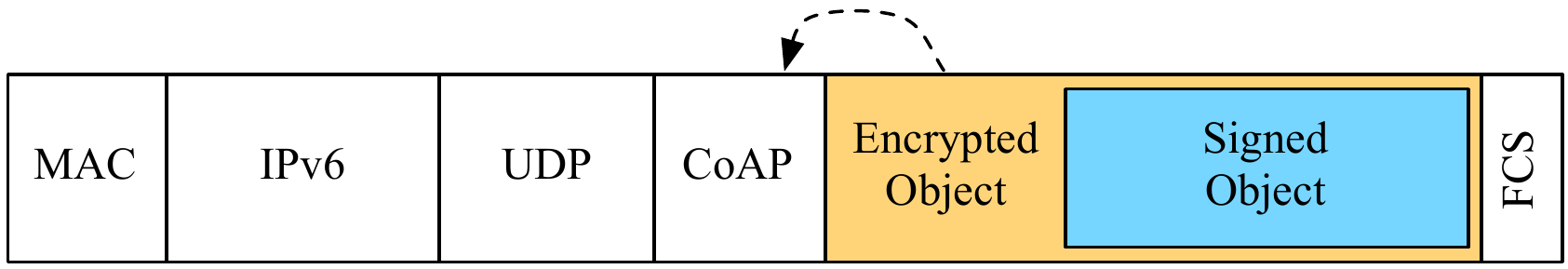}
\caption{Secured objects within the IoT network stack. The arrow represents the binding
of the object encryption key with the underlying CoAP header. In this configuration, 
signed object is regarded as the payload of the encrypted object, and is therefore encrypted.}
\label{fig:netstack}
\end{figure}
Protecting against replay attacks requires a state between end-points, which
contradicts our goal to provide a stateless approach to application
security. However, we exploit the fact that CoAP has been designed to run over
UDP and so, detect duplicates using a 16-bit MessageID in the protocol
header. We use this variable as salt to derive the content encryption key, when
responding to the current request. Furthermore, in order to avoid derivation of an identical
content encryption key among different senders in a possible group, we bind the key to the identity of
the sender, as discussed by Keoh \emph{et al.} \cite{dtls-multicast-draft}. We use
the unique identifier in the sender certificate known to all communicating parties.

More precisely, symmetric content encryption key $k_j$ is derived from access
secret $S_j$, current MessageID, and sender's unique identifier as: 

\begin{equation}
k_j = f(S_j, MessageID, senderID),
\end{equation}

\noindent where $f()$ is a generic pseudo-random function. Note that we keep the
authenticity of the content intact and use symmetric encryption as a means to
fight the replay and provide access control. We illustrate this in Fig. \ref{fig:netstack}. 
The replayed content will be detected as the derived decryption key will be different from the original key.
%FR: Message ID is used for duplicate detection, not to counter replay attacks, is there any need to talk about that anyway?
%In case replay attack is performed at lower network layers, we rely on the CoAP mechanism to detect it by means of MessageID.

However, this approach is vulnerable to replay attacks with substantially
delayed content, \emph{i.e.} once the CoAP client/server looses the MessageID context with communicating parties. To fight this, we rely on updates of the access secret, provided by the key management scheme.

\subsubsection{Cipher negotiation}

%AD: rephrase - do not understand....
As discussed in Section \ref{producer-consumer-model}, there is no motivation in traditional downgrade attacks.
%AD: which problem?
The problem of cipher negotiation then becomes trivial and is equivalent to content type negotiation
in a client-server interaction. Recall that a client in a REST like application
transfer protocol such as HTTP or CoAP, expresses its interest in content with a
\textit{GET} request. The request contains an optional ``Accept'' header
carrying preferred content types. If capable, server responds with the content supported by the client. Therefore, to solve the interoperability issues, we require an additional accept option carrying supported ciphers.

\subsection{Cryptographic Overhead}
\label{crypto} 

OSCAR ensures authenticity and integrity of the content by leveraging digital
signatures, which may seem surprising as we target constrained devices with
limited CPU resources. However, the use of public key operations at the level of
the content allows us to decouple the server-side cryptographic overhead from
network communication: constrained servers are able to update their
resource representations whenever it suits their schedule (take for example
energy harvested devices) and more importantly, while the radio transceiver is
turned off. The burden of digital signature verification is then put on clients,
as they should not consume the information before verifying its authenticity.

%AD: develop ECC acronym.
%MV 
The approach may seem surprising, but the evaluation results in Section~\ref{evaluation} suggest that Elliptic curve cryptography (ECC) public key operations
are actually less expensive than performing the pre-shared key DTLS handshake with every client, even in M2M scenarios, where both servers and clients are constrained.

Confidentiality of the content is ensured with symmetric encryption performed on a per-response basis by servers. For replay protection, OSCAR requires an additional per-response key derivation with typically lightweight cryptographic pseudo-random functions.

 \section{Security Considerations}
 \label{security}
 \subsubsection*{Denial of Service}
 %Separation of security concerns is the key to protecting applications running on a large number of highly constrained hosts.
 %[Extended version]

% Alongside with embedded devices, technological development and Moore's law
% apply to traditional consumer electronics used to access the Internet as well.
% As a consequence, it is reasonable to expect that low-cost sensing and
% actuating devices will always be orders of magnitude less gifted in terms of
% processing, memory and particularly energy supply. Thus, it is hard to
% envision that constrained devices will even in the future be fully capable of
% standing against resource rich counterparts in the Internet to fight energy
% exhaustion DoS attacks launched outside of the local network. We, thus,
% support discussions \cite{brachmann-e2e} that gateway(s) connecting the
% constrained networks to the rest of the Internet should provide means to fight
% DoS and control what packets are to be forwarded into the network. 
 
 %This could be done either by packet inspection and filtering, intrusion detection algorithms and/or by authentication and access control at the network layer with end hosts. 
 
OSCAR takes a non-traditional approach to fight Denial of Service. It builds upon the
assumption that typical IoT resource representations are small in size
(individual measurements of physical quantities, actuator state changes) and
directly responds to requests with access-protected resource
representations. Moreover, it does not keep any state between communicating
entities, which we find particularly important to fight memory exhaustion
attacks. Note also that since server-side digital signing operations are done
offline, the intensity of incoming traffic is not correlated with asymmetric cryptographic overhead. 
 %Nevertheless, any E2E security solution should ensure strong security properties to applications running on top. In that sense, we discuss security levels provided by OSCAR.

 \subsubsection*{Confidentiality}
As content encryption keys are derived from access secrets, OSCAR provides confidentiality within the resource access right group. Actual security properties are dependent on the encryption algorithm used. Note that an adversary able to compromise the Authorization Servers, may only obtain eavesdropping capabilities---E2E integrity and authenticity properties are preserved.

%AD: trust in terms of confidentiality?
% MV, yes, who do you trust to see the information but not to modify it.
If the mutual trust among clients in terms of confidentiality is not desired,
OSCAR puts the burden on the key management scheme running on Authorization
Servers. One such example would be the use of a recently proposed batch-based
group key management protocol \cite{veltri-batch}, where clients would be given
cryptographic material corresponding to descendants in the binary tree of the
actual access secret on a server. However, this would require additional signaling of the supported access secret in the GET request.

 \subsubsection*{Replay Protection}
OSCAR protects from replay at the level of the content by using an encryption
key that is a function of the MessageID from the underlying CoAP header. The detection of replay attacks performed at lower network layers depends on the CoAP duplicate detection mechanism. 
However, it is important to stress that the current CoAP draft, as is, would not provide robust protection in security terms. Therefore, successful coupling of OSCAR with CoAP would require additional clarifications and specifications to the duplicate detection mechanism.

Another concern with respect to the replay attack is a malicious adversary
within the resource access right group in case of asynchronous traffic. Such an
adversary is able to asynchronously inject old resource representations making
other members of the group believe they are fresh (if within the content itself,
there is no means allowing the detection of an old reading/command, \emph{i.e.}
a timestamp). Protection against such adversary would require the use of a key
%AD: differing -> different, OK?
management scheme that would provide different access secret cryptographic
material on the constrained server and individual clients, as discussed above.

 \section{Performance Evaluation}
\label{evaluation}
We have implemented an object security software library tailored for constrained
devices and the Contiki operating system that builds upon the open source
implementation of ECC cryptographic primitives---TinyECC (ContikiECC). We use
AES-CCM* as the symmetric encryption algorithm. The library supports creation,
parsing, and verification of ``encrypted'' and ``signed'' object
types. A certificate is then just a particular type of a ``signed object'' with a
pre-defined format.   Objects can be nested within each other to support 
%AD: Fig. 2 is not a use case.
%MV changed to netstack figure, it can pass as a use-case when it comes to object nesting
the use
case illustrated in Fig.~\ref{fig:netstack}. We have coupled the
object security library with Erbium CoAP, a default CoAP implementation for
Contiki (version 07) to add cipher suite negotiation capabilities (cf. Section
\ref{coupling-with-coap}). %The object security library is provided as
%open-source\footnote{The object security library for Contiki is available on request, contact Malisa.Vucinic@imag.fr to obtain a copy.}. 

\begin{figure}[htbp]
\centering
\includegraphics[width=0.9\columnwidth]{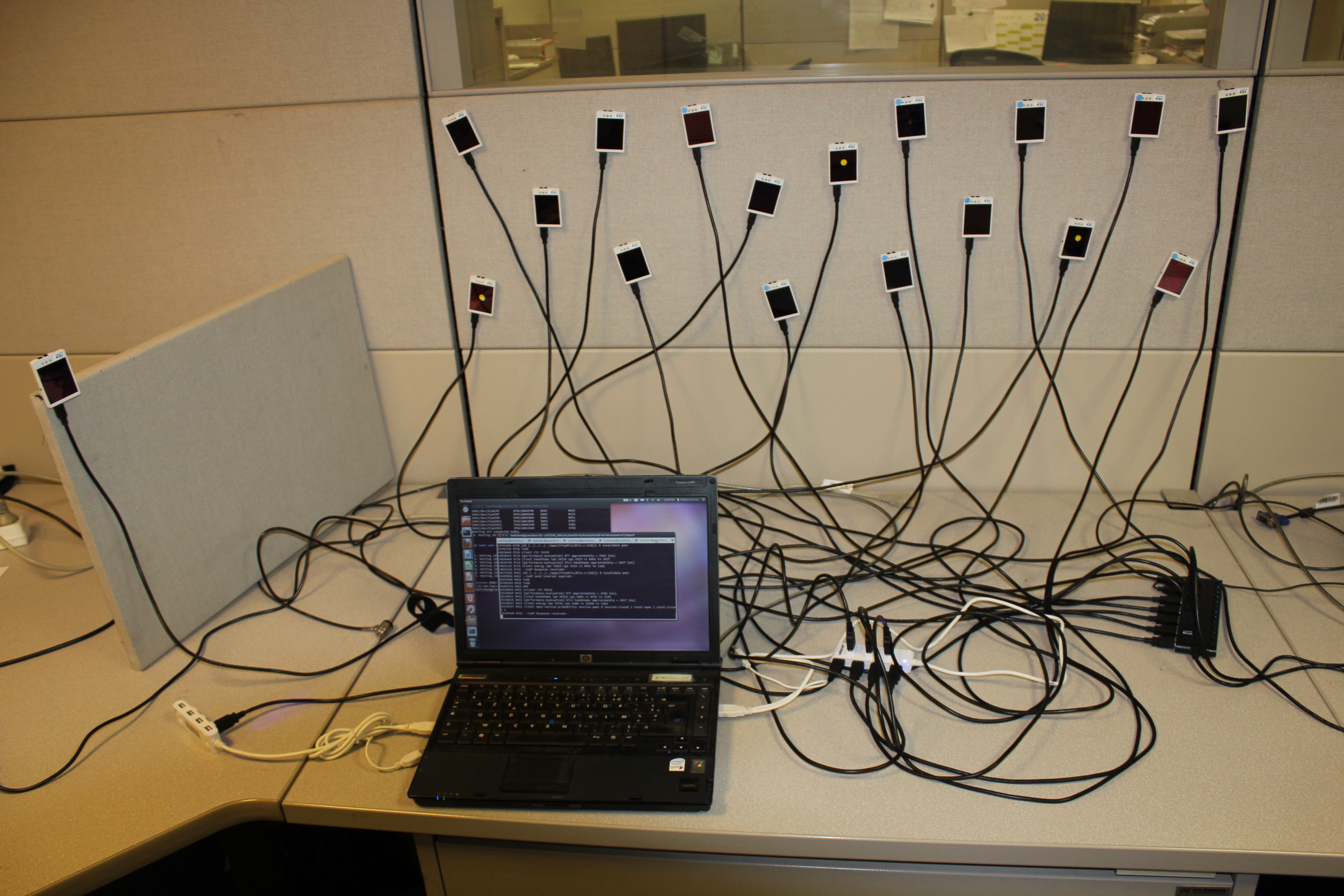}
\caption{Testbench with 18 energy-harvested ST GreenNet nodes in Crolles,
  France. 16 nodes are CoAP clients and one node is the CoAP server. The node on
  the far left is the PAN coordinator in the 802.15.4 beacon-enabled
  network. Nodes are connected to USB for the collection of experiment traces.
%Nodes are connected with USB for collection of experiment traces.
}
\label{testbench}
%\vspace{-3mm}
\end{figure}
Note that potential nesting of signed objects enables many additional features
that may be very useful for IoT use cases. For instance, a network gateway could add a global timestamp or location information to a signed object coming from a constrained node (constrained devices often do not have this information locally). 

%AD: ECDSA - def. ?
We evaluate two important aspects of OSCAR: 1) Elliptic Curve Digital Signature
Algorithm (ECDSA) computation overhead on constrained devices, 2)~scalability in M2M communication scenarios. Evaluations are performed for two hardware platforms that catch peculiarities of two generations of IoT devices:
\begin{itemize}
\item WiSMote platform based on 16-bit MSP430 (series 5) MCU with 16 KB of RAM
  and 802.15.4-compatible CC2520 radio transceiver. WiSMote related results are
  obtained using the instruction level MSP430 emulator MSPSim and the Contiki
  simulator Cooja, as we did not have enough real platforms needed for our
  experiments. However, we have confronted emulated measurements of ECDSA
  overhead in Cooja with those obtained on real WiSMote hardware and we have measured a maximum error of 2.67\%. 
\item ST GreenNet tag, an energy-harvested prototype platform from STMicroelectronics (ST) based on an ultra low power 32-bit ARM~Cortex-M3 MCU (STM32L) with 32 KB of RAM and an 802.15.4 radio transceiver. ST GreenNet results are obtained from real hardware.
\end{itemize} 

To eliminate the effect of a variable CPU frequency on results, we have configured both platforms at 21.3 MHz. MSP430 series 5 may be configured up to 24 MHz, while the STM32L can go up to 32 MHz. Both computation time (inversely proportional) and CPU energy consumption (directly proportional) are linearly dependent on frequency.

We estimate energy consumption using Energest, a Contiki per-component profiling
tool. Energest effectively measures the time spent by  different components on a
platform in a given state (for instance, the time CPU spent in active or low
power mode; radio transceiver in RX or TX). These values are converted to energy by
multiplying with the constant operating voltage (we used 2.8~V) and the current draw values from appropriate data sheets. %We ported Energest to the ST GreenNet platform and verified with oscilloscope its precision. We measured a maximal error of 2.439 \%.

%We ported Energest to the ST GreenNet platform and verified with oscilloscope
%its precision. We measured a maximal error of 2.439 \%.

\subsection{ECDSA Computation Overhead}
\begin{figure}[htp]

\noindent
\centering
\subfigure[Computation time.]{
\includegraphics[width=0.48\columnwidth]{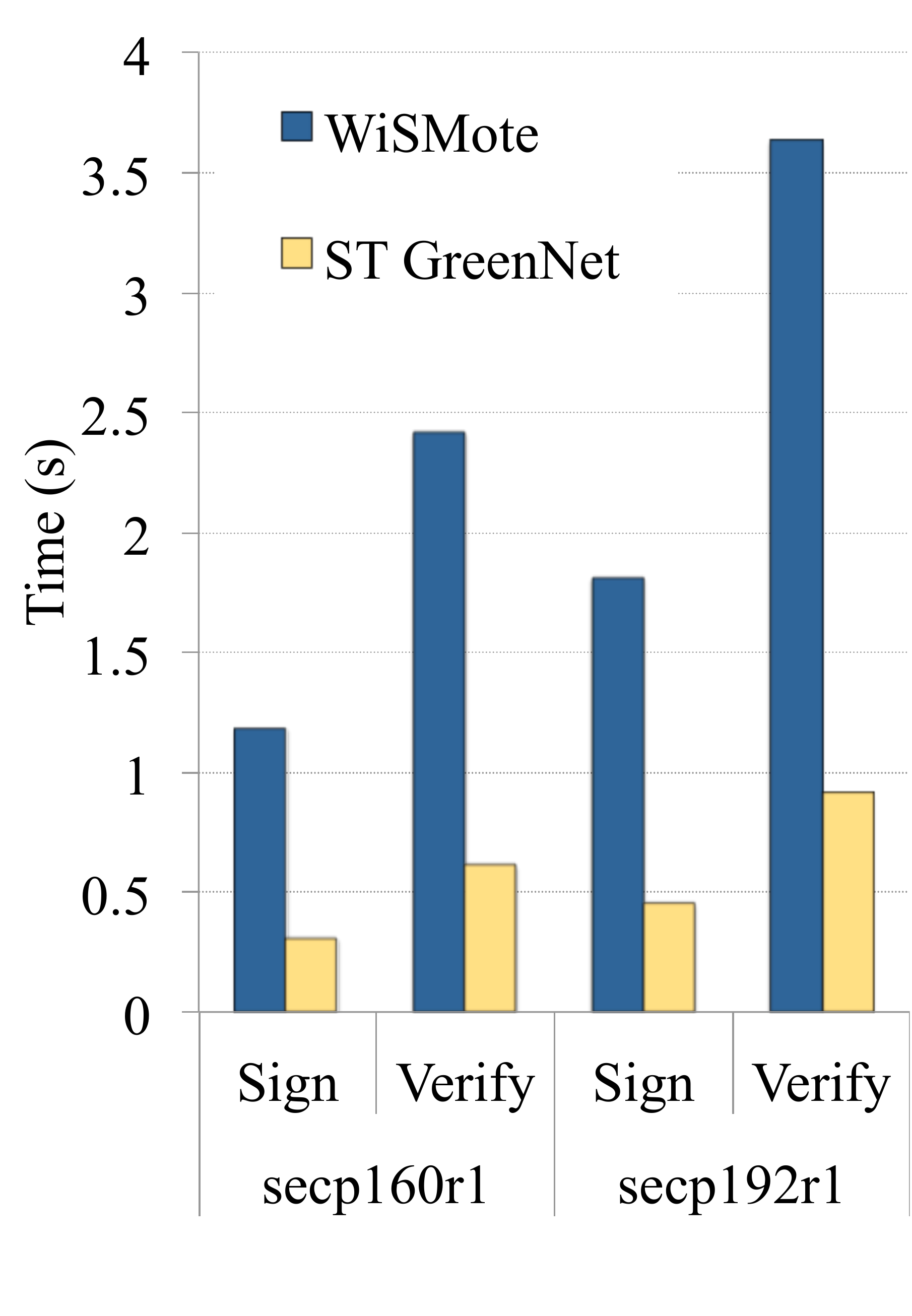}
\label{ecdsa-time}
}
\hspace{-0.37cm}
\subfigure[Energy consumption at 2.8V.]{
\includegraphics[width=0.48\columnwidth]{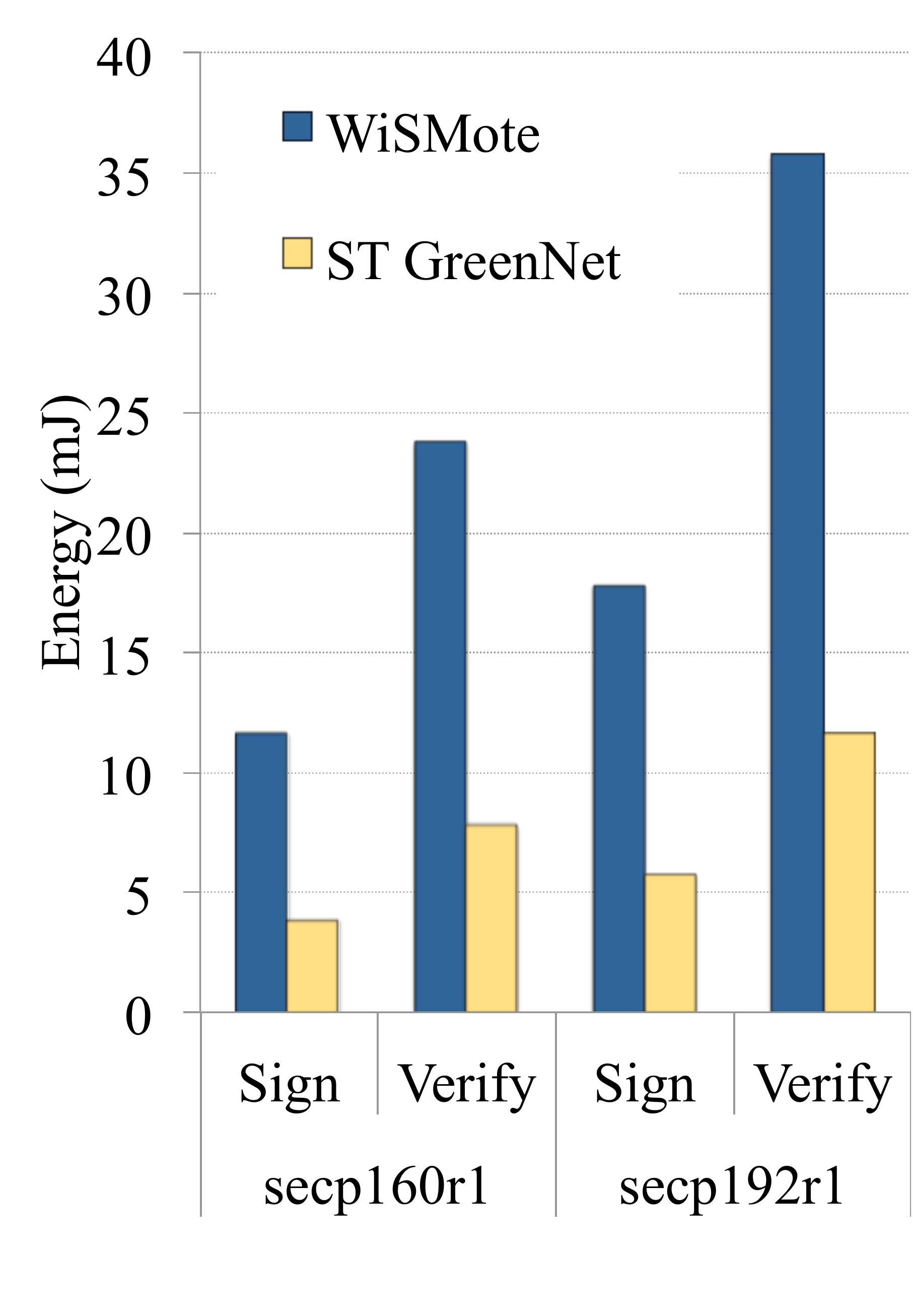}
\label{ecdsa-energy}
}
\caption{ECDSA computation and energy benchmarks at 21.3 MHz for
  16-bit~(WiSMote) and 32-bit (ST GreenNet) hardware platforms. We use TinyECC
  (ContikiECC) open source library. Message size is of 25 bytes.}
\end{figure}

Figs. \ref{ecdsa-time} and \ref{ecdsa-energy} present computation and energy
benchmarks of the ECDSA primitives (secp160r1 and secp192r1 elliptic curves) on
WiSMote and ST GreenNet platforms. We can see that the use of a 32-bit MCU
reduces the computation time by a factor of 4, which translates into a reduction
in the consumed energy by a factor of 3.084 (as the 32-bit STM32L consumes 29.7\% more than 16-bit MSP430 in active mode). 

Figs. \ref{ecdsa-time} and \ref{ecdsa-energy} strongly support our initial
design assumption on processing capabilities (cf. Section
\ref{trends}). Whatsoever, we expect that further advancements in
nano-technology will additionally reduce the energy computation cost for low power MCUs.

Still, computation overheads ranging from 0.3 to 0.9 seconds for the 32-bit
platform and from 1.18 to 3.63 seconds for the 16-bit platform, at the first
sight seem like a huge price to pay. In fact, Hummen \emph{et al.} argue that
for this reason, the number of public key operations should be minimized during the security handshake  \cite{hummen-certificate}. OSCAR, however, compensates for this overhead by removing the radio energy cost of the security handshake with every client.

In the following section, our goal is to determine if OSCAR and the heavy use of ECC public key primitives outperform a connection-oriented approach with DTLS that uses only lightweight symmetric key operations during the handshake.
%That is, we compare  the overhead introduced by means of public key operations and intensive CPU usage with that of increased radio traffic due to the DTLS handshake.

\subsection{Scalability}

We study scalability as a function of the ratio between the total number of
clients and a maximum number of open DTLS sessions at a constrained server (due
to memory limitations, constrained servers have a limited number of DTLS session
``slots''). 
We have followed the guideline on practical issues with DTLS
(Section 2.1 \cite{dtls-practical-issues}) and extended the TinyDTLS
implementation with the Least Recently Used (LRU) session closure algorithm. The
server immediately releases memory and sends a closing alert to the LRU session
as soon as a new client has demonstrated good intentions by retransmitting the
stateless cookie in the ClientHello message (recall the DTLS
handshake). Therefore, the handshake with the new client proceeds immediately. 
%AD: rephrase... - don't understand this phrase.
%Other than the LRU closure policy when all slots are occupied, sessions are indefinitely kept open.
Clients keep their sessions open as long as possible, \emph{i.e.} until they receive the closing alert
from the server.

The maximum number of DTLS session slots is dependent on platform memory
capabilities and actual application memory requirements. With the full IPv6
networking stack of Contiki and a simple application for evaluation purposes, we
were able to have a maximum of 3 session slots on WiSMote (TinyDTLS
implementation). However, as stressed out, this number should not be generalized
as it depends on the implementation specifics of an application and the
operating system. We have used the same number of slots on the ST GreenNet
platform as well to have comparable results. Although a higher number of slots
would be available due to a larger memory, the results in terms of the ratio do
not loose generality.

\begin{figure*}[htp]
\noindent
\centering
\subfigure[WiSMote]{
\includegraphics[width=1\columnwidth]{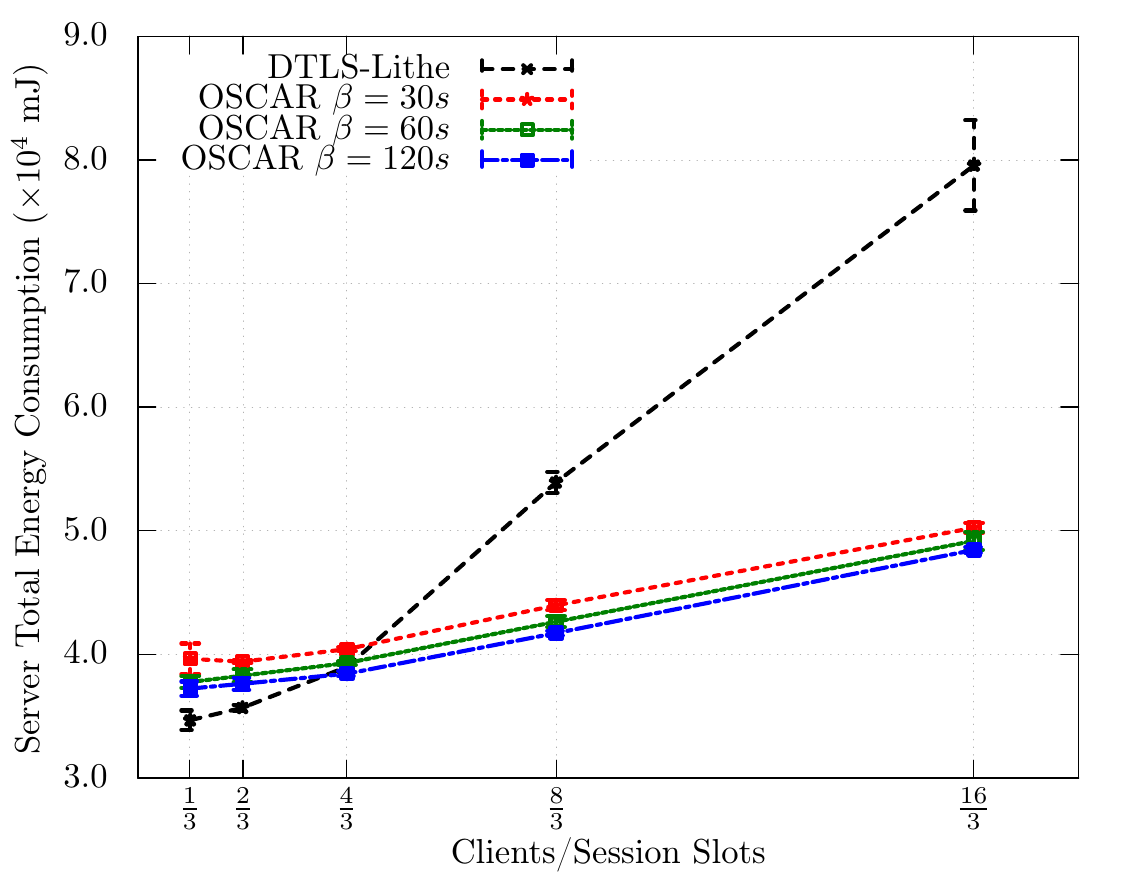}
\label{server-total-wismote}
}
\hspace{-0.37cm}
\subfigure[ST GreenNet]{
\includegraphics[width=1\columnwidth]{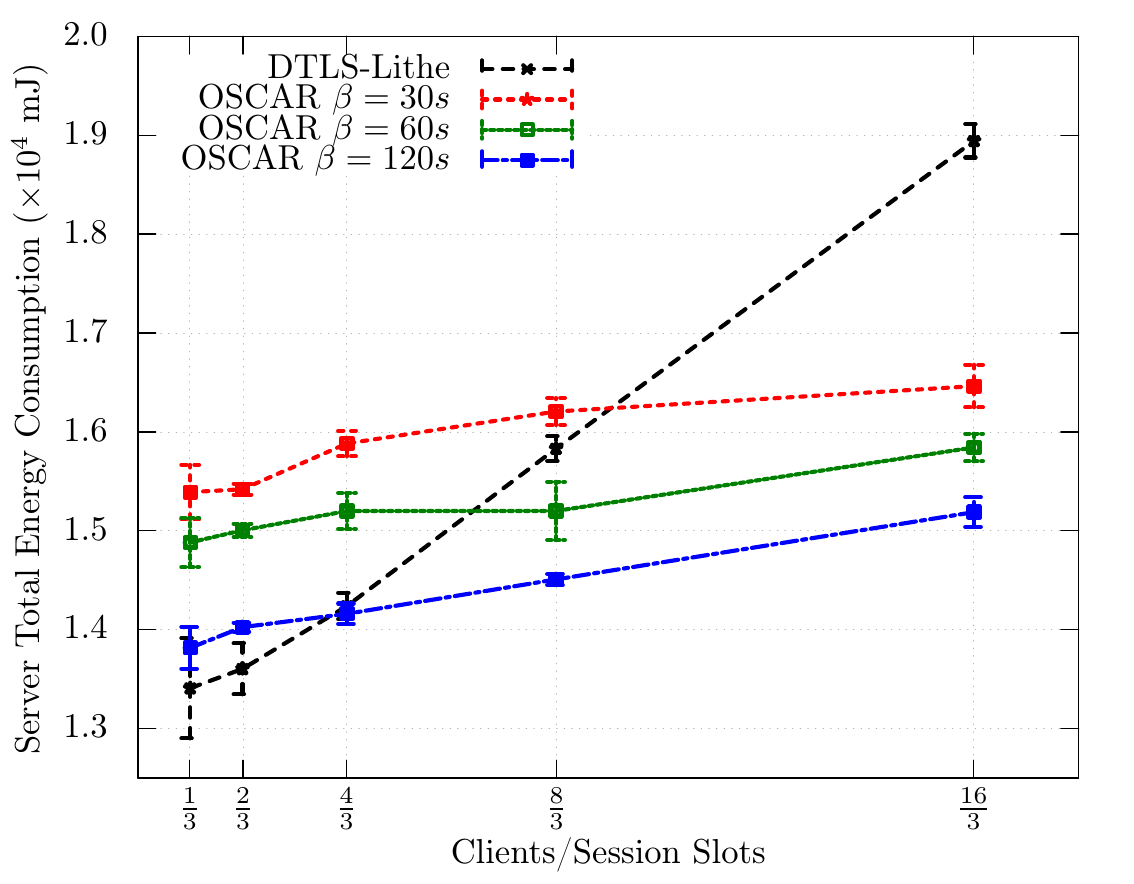}
\label{server-total-greennet}
}

\caption{Constrained server total energy consumption over 3 hours.}
\end{figure*} 

Table \ref{tab:config} shows the configuration of the two platforms. Note that
we use two different Radio Duty Cycling (RDC) protocols.
Thus, we demonstrate the performance of OSCAR and DTLS running on top of asynchronous (X-MAC)
and synchronous (beacon-enabled IEEE 802.15.4) RDC protocols, which covers the vast majority of IoT use cases.
We set the Beacon Interval of beacon-enabled 802.15.4 to 122.88 ms to have
comparable delays with X-MAC (default channel check rate of 8~Hz). Simulations
in Cooja were performed for a star network topology with the CoAP server being
the central node (radio neighbor for each client and preferred parent in the RPL
DODAG). Note that due to the specifics of beacon-enabled 802.15.4, one node in the
network has a mere role of being the PAN coordinator and transmitting periodic
beacons. Other nodes in the network associate with it (L2 operation), which
effectively introduces an extra hop between CoAP clients and the CoAP server, in
respect to the network evaluated in Cooja (cf. Fig. \ref{testbench}). %However,
%our goal is not to compare the two platforms or RDC layers, but rather to
%demonstrate the performance of OSCAR and DTLS running on top for two different generations of IoT devices.

In both scenarios, nodes run a typical IPv6 networking stack over 802.15.4
(CoAP, UDP, IPv6, 6LoWPAN). We use a recent 6LoWPAN compression scheme of DTLS
named Lithe \cite{lithe} to maximize its performance. In the case of ST GreenNet platform,
we use the proprietary implementation of beacon-enabled 802.15.4, as in our previous work
 \cite{l2energycost, rpl-topologyconstruction}. CoAP clients send a single
GET request for a resource on the server according to the exponential
distribution with a mean of 0.5~requests per minute. If the DTLS session is
found open, the request is sent directly without waiting for the handshake to
complete. If not, the client first performs a DTLS handshake with the server. Responses contain an abstract resource representation with 25 byte length. In case of OSCAR, this representation is transferred as the appropriate encrypted and signed object type.

An important aspect for performance evaluation of OSCAR is the server-side
resource signing load. We define parameter $\beta$ as the mean resigning
interval such that $\beta=t/N$, where $N$ is the total number of secured
resources on the server and $t$ is the average resource update time (for
instance, updates of temperature, pressure, CO$_2$, etc.). We evaluate OSCAR for
$\beta$ values of 30, 60, and 120 seconds, to account for use cases where high, medium, or low signing load is needed. 

In the case of OSCAR, we use pre-shared access secrets and certificates to
decrypt and verify encrypted and signed objects. Similarly to the work of Hummen
\emph{et al.} \cite{hummen-certificate}, we use the secp160r1 elliptic
curve. Objects are encrypted using the AES-CCM* algorithm. Similar assumptions
apply to DTLS as well: it uses the TLS\_PSK\_WITH\_AES\_128\_CCM\_8 pre-shared key based cipher suite. As a consequence, DTLS only uses symmetric key operations during the handshake.

We have run experiments/emulations over 3 hours and plotted 5 run averages with 95\% confidence intervals.

\begin{table}[htbp]
\caption{Experiment setup.\label{tab:params}
%\vspace{-3mm}
}
\centering
\subtable[WiSMote]{
 \begin{tabular}{cl}
        \toprule
         {{Radio Duty Cycling}} & {X-MAC}  \\
        %\hline
       {{Channel Check Rate} (Hz)} & {8}   \\
        %\hline
        {Channel Model} & {Unit Disk Graph}  \\
        \bottomrule
      \end{tabular}     
%      \label{fig:1dag}
}
\subtable[ST GreenNet]{
 \begin{tabular}{cl}
        \toprule
         {{Radio Duty Cycling}} & {beacon-enabled 802.15.4}  \\
        %\hline
        {{Beacon Interval} (ms)} & {122.88}   \\
        %\hline
        {Superframe Duration (ms)} & {15.36}  \\
        \bottomrule
      \end{tabular}     
%      \label{fig:1dag}
}
\label{tab:config}
\end{table}

Figs. \ref{server-total-wismote} and \ref{server-total-greennet} show the effect
of the reduced radio traffic generated by OSCAR on energy consumption. For
medium intensity signing load ($\beta = 60 s$), in case of WiSMote, OSCAR
crosses the energy performance of compressed DTLS when the client/session slot
ratio is approximately 1.3. In case of the ST GreenNet platform, the crossing
is increased to approximately 2.15 due to the use of a new generation
(prototype) radio with lower consumption. It is important to stress that the
exact crossings depend on the consumption characteristics of the MCU and the radio transceiver, and our results are therefore particular for the two evaluated platforms. However, the MCU/radio transceiver combinations on the evaluated platforms are very representative---16-bit CPU and an old generation radio (WiSMote) and a powerful 32-bit CPU with prototype low consumption radio transceiver (ST GreenNet) allowing us to generalize the crossings between the two.

\begin{figure}[htbp]
\centering
\includegraphics[width=1\columnwidth]{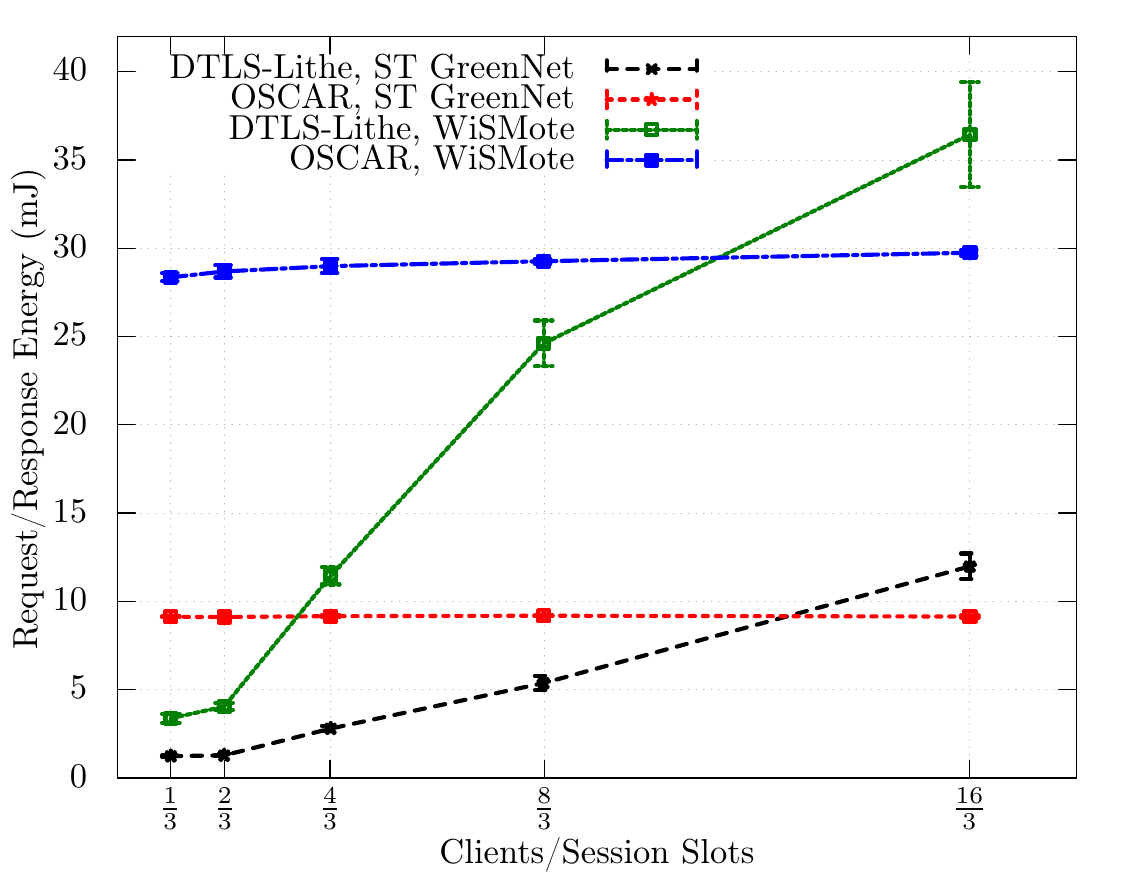}
\caption{Client energy consumption per CoAP request-response. It includes a possible DTLS handshake.}
\label{client-pr-energy}
%\vspace{-3mm}
\end{figure}

Although our initial design goal was to relieve constrained servers from radio
traffic and to place burden on clients, we can notice in
Fig. \ref{client-pr-energy} that even for moderate (in IoT terms) client/session
slot ratio (WiSMote 3.7, ST GreenNet 4.17), constant ECDSA verification results
in better performance than using the compressed DTLS approach. Note that in our
evaluations, we use constrained clients as well thus accounting for the
worst case. In IoT use cases, it is expected that a significant part of clients will be more powerful devices such as smartphones, tablets, laptops, or powerful cloud servers.

\begin{figure}[htbp]
\centering
\includegraphics[width=1\columnwidth]{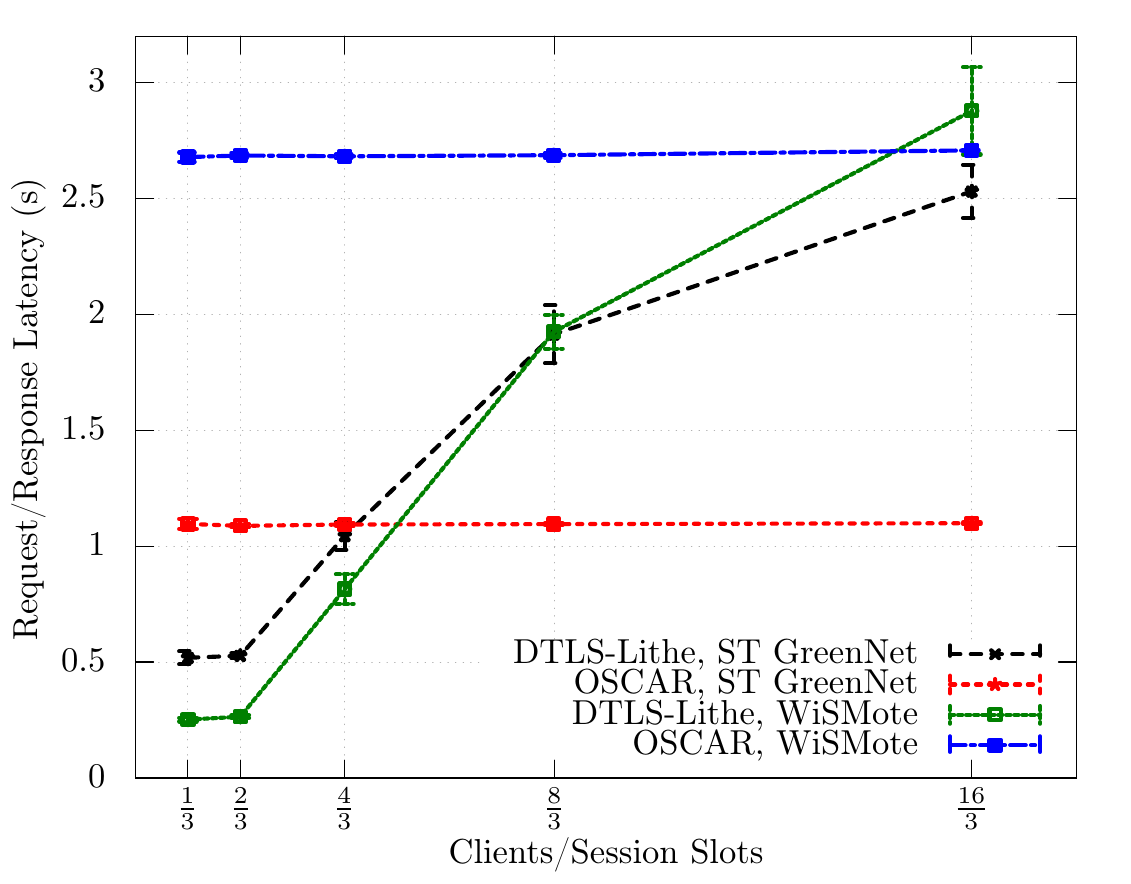}
\caption{Request-response latency. It includes a possible DTLS handshake.}
\label{client-rtt}
%\vspace{-5mm}
\end{figure}

Finally, we evaluate the request-response latency in
Fig. \ref{client-rtt}. As we can see, MCU computation capabilities greatly
affect the result of OSCAR as most of the latency comes from ECDSA
verification. In the ST GreenNet deployment, we have observed an increased
number of failed DTLS handshakes for the largest evaluated network with 16
clients due to the stochastic nature of radio links. Note that DTLS curves
exponentially increase with the number of clients, but are expected to saturate
for denser networks. The exact saturation point depends on the configuration of
the DTLS retransmission mechanism (we have used the default retransmission timeout of 2 seconds). 

\section{Securing the Internet of Things} % Maybe add state of the art in the title [???]
\label{relatedwork}

%Short paragraph summarizing different level approaches so far. State that 6LoWPAN does not provide any security services and that it relies on link layer security of 15.4 to protect the network. State that the main CoAP draft discusses the use of DTLS to provide security, while IPsec is also mentioned as an alternative in a separate draft. Also, that it mentions ``object security'' which is our approach.
Research and standardization efforts around secure IoT follow the TCP/IP architectural model by having security features on one (or more) of the layers in the protocol stack. Accordingly, we survey the state of the art.
\subsection{End-to-End Security at the Network Layer}
\label{ipsec}
% Survey the work supporting IPsec to provide E2E security

% Describe practical problems: 
% - not yet specified for compression by 6LoWPAN - check if any work is ongoing
% - it's in the kernel so it is not so easy to enable it on the Internet side
% - give numbers of added overhead of the proposed 6lowpan compression from SICS 
% - discuss that it is generally used to secure VPNs [???] and that it is a heavy weight option. 

Ever since the efforts on integrating Wireless Sensor Networks with the Internet
have begun, the so-called blanket coverage at the network layer has been
considered a potential solution to provide end-to-end security services
\cite{minisec}.  The literature widely discussed the feasibility of porting the
IPsec protocol suite to smart objects \cite{granjal2008ipsec,
  roman2009integrating, granjal2010secure, raza2011securing, raza2012journal,
  granjal2012effectiveness}. The authors mostly evaluated the processing overhead and energy requirements of different cryptographic suites used by IPsec, but also the memory footprints and system response time \cite{raza2011securing, raza2012journal}. Even though it was initially considered too heavy for constrained environments, these results led to the common conclusion that a lightweight version of IPsec is a feasible option.

In the Internet, IPsec mostly secures Virtual Private Networks (VPN). Being at
the Network layer, it is perfectly suited for such applications where
``blanket'' coverage is actually desirable (enterprise networks for
example). However, as it resides in the Operating System kernel, it is
impractical for typical IoT applications. The requirement that an end user needs
to configure the host Operating System and IPsec for securing communication with
smart objects would probably result in questionable security practices. 
%AD: translation?
%MV changed to mapping
Moreover, integrity at the Network layer would prevent any protocol mappings. Namely, as the IP payload is being authenticated, there would be no way of performing HTTP/CoAP mapping at the network gateway. CoAP, however, has been designed from the very beginning to facilitate this for legacy hosts in the Internet.
%Briefly add numbers on overhead ; Add figure common with DTLS on available payload size with proposed compression schemes later.

\subsection{End-to-End Security at the Transport Layer}
\label{transport-layer-sec}
% State that it is the most intuitive approach knowing the wide spread of use in current Internet.
% Mention Sizzle , SSNAIL and other work done before CoAP.

Impracticality of IPsec has been overcome in the Internet by introducing the
security services just below the application layer, in the form of TLS/SSL. The
wide and successful use of this model in the Web has also suggested its use in
IoT. Indeed, the first proposal on using SSL for smart objects, nicknamed
Sizzle, came in 2005 from Sun Microsystems \cite{sizzle}. The authors evaluated
the HTTPS stack that leverages assembly optimized implementation of  ECC as a public key algorithm. At the time of the
publication, however, there was no common agreement on the transport protocol to
use. Consequently, the authors implemented their own reliable transport
protocol. SNAIL \cite{snail} complemented this work by introducing SSL on an all
IP architecture, leveraging the 6LoWPAN adaptation efforts done in the
meantime. Together with the introduction of IP to the embedded world came the
dilemma whether TCP is suited or not, due to its connection establishment overhead, poor performance in case of lossy networks and short term connections. For this reason, latest standardization efforts \cite{coap-draft} assume User Datagram Protocol (UDP) at the transport layer, leaving reliability as an option to the application.

Unreliable transport and possible out of order delivery make TLS as is, an
improper candidate for IoT. For the reason of securing application level
protocols running over UDP in the Internet, such as Session Initiation Protocol
(SIP), Real Time Protocol (RTP), or Media Gateway Control Protocol (MGCP), TLS
has already been extended to Datagram TLS (DTLS) \cite{dtls-rfc, dtls}, which
introduced additional 8 bytes of per datagram overhead in the form of the
%AD: from? from TCP?
sequence and epoch numbers that were implicitly known with the reliable
transport. 

% Survey the work supporting (D)TLS.

As a straightforward and standardized parallel to the successful model in the
Internet, DTLS has attracted attention of the research community around the Internet of Things  \cite{raza20126lowpan, brachmann-e2e, kothmayr2012dtls, granjal2012effectiveness, lithe, hummen-certificate}.  It is interesting to note, however, that apart from the known advantage of using an already standardized protocol, no argument has been given on actual applicability of DTLS for IoT.
Kothmayr \emph{et al.} \cite{kothmayr2012dtls} discussed the necessity of authenticating both the client and the server during the DTLS handshake, but their experimental results show significant completion delays, ranging from 2 to 6.5 seconds. Granjal \emph{et al.} \cite{granjal2012effectiveness} performed a comparative study on memory footprints, computational time, and required energy between IPsec protocols and DTLS, using different cryptographic suites. These results showed similar performance of the two approaches, except in the case when DTLS is additionally used to exchange keys with 
the Elliptic curve Diffie-Hellman exchange.

Recognizing the excessive overhead of the DTLS handshake, Hummen \emph{et al.}
\cite{hummen-certificate} proposed different techniques to lower its impact on
constrained devices---certificate pre-validation and handshake delegation to the
network gateway. On the other hand, Raza \emph{et al.} tackled the same problem
by proposing a 6LoWPAN DTLS compression scheme \cite{raza20126lowpan} that
reduces per datagram overhead. This work has lately been integrated with CoAP
and released in the open source form \cite{lithe}.

% Specification proposal for 6LoWPAN DTLS compression by Raza \emph{et al.} \cite{raza20126lowpan} reduces the record layer overhead by 5 bytes, resulting in at least 8 signalization bytes added to each record. % Note that the variable size message authentication code makes part of the payload of the DTLS record layer. 

A significant drawback of using DTLS to secure IoT is its incompatibility with
multicast traffic. As stated by its designers~\cite{dtls}, DTLS targets
typical connection-oriented client-server architectures. While some of the IoT
envisioned applications could loosely undergo this assumption, the majority
cannot (cf. Section \ref{internet-trust-model}). In fact, group communication support is one of the main features why CoAP protocol is being standardized at all \cite{coap-draft}. %Within IETF, there has been a proposal on a possible DTLS based security architecture encompassing additional security protocols for authentication and key management (PANA and AMIKEY) \cite{multicast-dtls}, in order to provide network access and multicast support. Redundancy in providing security services with such approaches is obvious, as DTLS is being concurrently used. Not to mention the additional implementation complexity for already constrained devices that would be introduced. The use of similar ad-hoc, complex and patched solutions would probably result in group communications not being secured at all. 

%Contrary with this approach, in this paper we present a direction how the well-studied properties of DTLS can be fully leveraged while  

% Describe problems of having DTLS in the IoT, discussed in the literature:
% - Not yet specified for compression by 6lowpan - check if any work is ongoing
% - No support for multicast; Cite the draft proposing a separate DTLS session with each member of the multicast group in order to exchange the multicast key.
% - Give numbers of added overhead of the proposed 6lowpan compression scheme from SICS and what would actually be left for the application to use

% - Discuss the problem if CoAP/HTTP protocol translation is performed at the border router and the propositions to map DTLS to TLS, etc... Def. add figure on this
% [MV] Following paragraph may be taken out to save space
Additional concern raised by the straightforward, point-to-point use of DTLS is
incompatibility with scenarios where the end-host in the Internet only supports HTTP/TLS. As stated previously, CoAP has been designed from the beginning to provide easy mapping to HTTP. Brachmann \emph{et al.} \cite{brachmann-e2e} discussed a possible DTLS/TLS mapping done at the gateway that preserves E2E security. While verifying integrity at the transport layer, however, it is impossible to perform the CoAP/HTTP mapping at the application layer.

\subsection{Object Security Approaches}
Although the concept of object security, \emph{i.e.} placing security within the
application payload, has been discussed as an option \cite{coap-draft,
  cirani2013enforcing} the related work in the literature leverages its benefits to provide fine grained access control with an assertion-based authorization framework \cite{authorization-framework}. %Our work primarily tackles the problem of E2E security and the capability-based access control is solely a means to provide communication confidentiality.
We jointly address the problems of E2E security and authorization for IoT and use the
capability-based access control solely as a means to provide communication confidentiality. 
The work of Seitz \emph{et al.} \cite{authorization-framework} is complementary to ours and the
two approaches can be merged to cover the use cases where simply responding with an
access-protected resource representation is undesired.

\subsection{Standardization Efforts}
Recent IETF efforts are directed towards profiling DTLS specifically for
constrained devices (DICE working group). Current proposals aim at adding
multicast support to DTLS by reusing the record layer and relying on an
independent group key management protocol \cite{dtls-multicast-draft}. In
essence, the core (D)TLS design assumption (point-to-point communication) is
being revisited to make it fit better the IoT requirements.

Authorization and authentication challenges for constrained environments are being tackled
separately within the ACE working group. Requirements that are discussed by ACE,
however, seem to be contradictory with the initial choice of DTLS as a security protocol,
 particularly when it comes to proxies and caching. OSCAR bridges this gap and
 jointly tackles the problems
 of E2E security and authorization, while keeping full compatibility with the plain DTLS approach. 

On the other hand, 6TiSCH working group of IETF designs a security architecture
to enable bootstrapping of IEEE 802.15.4 nodes. The main challenges include initial network access 
and the setup of L2 keys using existing IP protocols.

Finally, it is important to note that different standards specifying the object security format already exist or 
are under standardization (Cryptographic Message Syntax---CMS, JSON Object Signing and 
Encryption---JOSE), but their adaptation for constrained devices is required. 
% - State that adding connection-oriented security protocol on top of connection-less transport is just wrong [???]
%In any case, most of the discussed problems come from the simple fact that we are trying, no matter what, to incorporate a successful server-client security model into an architecture with substantially different application requirements. We discuss this in Section~\ref{internet-trust-model}.

\section{Conclusion}
\label{conclusion}
Our work explores a novel approach to the problem of E2E security in IoT. It is
based on the concept of object security that introduces security
within the application payload. We consider separate confidentiality and
authenticity trust domains. Confidentiality is used as a means to provide
capability-based access control and a protection against eavesdropping during
the communication. We protect from replay attacks by coupling the content
encryption key with the duplicate detection mechanism of CoAP. 
Authenticity is tied to the host and the content encapsulated within
objects is digitally signed, which allows the trust in the information to
persist long after the actual communication has taken place. In turn, this
property enables local databases and caches to use the intrinsically secure
content. Moreover, leveraging the access right confidentiality domain and the
concept of object security, our proposal intrinsically supports multicast. We
take off the burden of a security handshake with every client from constrained
servers. Instead, we rely on secure communication channels with Authorization
Servers that are in charge of resource access right key
management. Cryptographic burden is then shifted to clients that need to perform
signature verifications for the content they are interested in.

We have demonstrated the feasibility of the concept by evaluating the proposal in
the M2M communication scenario where all parties are resource constrained. We
believe that billions of smart, but constrained objects, encompassing IoT are the
best argument for a scalable solution such as OSCAR. OSCAR is particularly
useful in Smart City deployments where energy constrained servers are expected to have
a  large number of clients. 
%Performing the security handshake with every client would put a significant burden on the servers in terms of energy consumption. 

We are aware of the fact that the work in this paper may not meet requirements
for each and every use case of the vast IoT domain. More specifically, the use
cases that require streaming
%. In fact, specific techniques that would maximize
%the security and applicability of OSCAR to other use cases
 are part of an
on-going work. 
%AD: already explained.
% Our goal in this paper was, rather, to present a system level
% approach to the problem of E2E security, to demonstrate feasibility, and discuss
% benefits in terms of flexibility that the concept of object security brings
% along. 

\section*{Acknowledgments}
We would like to thank Shahid Raza, Hossein Shafagh and Simon Duquennoy for releasing the
implementation of Lithe in open sourced form. Many thanks to Michel Courbon for performing tests on real 
WiSMote nodes and to Michel Favre for suggestions on porting the Energest benchmarking tool to ST 
GreenNet platform.
The work of F. Rousseau and A. Duda was partially supported by the
French National Research Agency (ANR) project project IRIS under contract
ANR-11-INFR-016 and the European Commission FP7 project CALIPSO under contract
288879. The work reflects only the authors views; the European Community is not
liable for any use that may be made of the information contained herein. 

%\nocite{*}
\bibliographystyle{IEEEtran} 
\bibliography{IEEEabrv,./Biblio/cloud,./Biblio/content_centric_networking,./Biblio/iot_security,./Biblio/rfcs_and_drafts}

\end{document}